\documentclass[11pt]{article}
\usepackage{mathptmx}
\usepackage{amsmath, amssymb}
\usepackage{graphicx}
\usepackage{booktabs}
\usepackage{url}
\usepackage{fullpage}
\usepackage{parskip}

\newcommand{\R}{\mathbb{R}}
\graphicspath{{./}}

\title{Fast Spammer Detection Using Structural Rank}

\author{Seungyeon Kim\\Georgia Institute of Technology\\{\tt seungyeon.kim@gatech.edu}
    \and Haesun Park\\Georgia Institute of Technology\\ {\tt hpark@cc.gatech.edu}
    \and Guy Lebanon\\Amazon\\{\tt glebanon@gmail.com}}
\date{}

\begin{document}
\maketitle

\begin{abstract}
\noindent
Comments for a product or a news article are rapidly growing and became a medium of measuring quality products or services. Consequently, spammers have been emerged in this area to bias them toward their favor. In this paper, we propose an efficient spammer detection method using structural rank of author specific term-document matrices. The use of structural rank was found effective and far faster than similar methods.
\end{abstract}

\section{Introduction}

Comments are widely growing in web: for a product, a news article, or a sport game. They are usually much shorter than main articles, but large in number with numerous authors. The rich set of responses from varieties of audience, comments, now are analyzed by lots of researchers who believe that they show important aspects of the main article.

Spammers also found the importance of the comments and started to abuse all kinds of comment system and overwhelmed other legitimate comments. They usually exploits the comment system with automatic programs to keep posting their propaganda.

The behavior of the spammers shows a critical point to detect them. Most of their contents are repeated under the same author name. There are sever researches \cite{Mishne2005a,Jindal2007,Lim2010,Fei2013} that compute content similarity to detect such spammers. They make use of language model \cite{Mishne2005a}, or set intersection similarity \cite{Broder1997,Jindal2007}, or average cosine similarity \cite{Lim2010,Fei2013}. In this paper, we will use much faster but an effective method: the structural rank of author specific term-document matrix to detect the spammers.

\section{Related Work}

Our work is close to social media spam detection as they usually deal with short documents with large number of authors. The approaches are slightly different from traditional spam detection which focuses emails or websites. \cite{Heymann2007} is a good survey of dealing with spam in social media.

Various content-based features were found effective detecting spams or spammers. \cite{Mishne2005a} used language models to detect spams in blog posts. Bag-of-anchors and bag-of-url were used in \cite{Kolari2006}. \cite{Cattuto2007} defined folksonomy which are tags co-occurancing in network neighbors to detect spammers. \cite{Lim2010,Fei2013} computed average all-pair cosine similarity of one specific author with the help of other features.

User networking behavior were also well studied in this area. \cite{Krause2008} make use of tagging behavior of a user, such as user concurrence with other spammers. \cite{Benevenuto2009} did similar approach categorizing users on Youtube into spammers, promoters, or legitimate users. \cite{Moh2010} did similar approach on Twitter. \cite{Lim2010} proposed a behavior model adding review score features testing such as its fairness with other features. \cite{Fei2013} additionally takes in to account trends of the review (called burstness of review). Graph similarity based detection was used in \cite{Wang2011}.

\section{Structural Rank}\label{sec:structural-rank}

The structural rank is the maximum rank of all possible matrices of the same non-zero pattern. Since it only considers the non-zero patterns, we can make use of bipartite graph traverse algorithms for an efficient computation instead of traditional methods for numerical (or theoretical) rank of the matrix. Computing the numerical rank takes $O(m n^2)$ with a matrix $A\in\R^{m \times n}, m \ge n$ by computing only singular values using SVD.

There are efficient algorithms computing structural rank \cite{Pothen1990,Duff1981}. The worst case time complexity was shown $O(\tau n)$ where $\tau$ is the number of non-zero entries of the matrix; however, \cite{Duff1981} also showed it will run $O(\tau + n)$ in most practices.

The computation benefit is easily noticeable in sparse matrices. In sparse matrices, we know number of non-zero entries are much smaller than the size of the matrix: $\tau \ll mn$. Consequently, $O(\tau n) \ll O(m n^2)$. Moreover, the practical bound $O(\tau + n)$ is obviously smaller than $O(m n^2)$. In section~\ref{sec:computation}, we will empirically compare the computation speed.

\section{Spammer Detection}\label{sec:method}

\subsection{Computing Content Similarity of a Set of Documents}

We assume spammers will keep posting similar contents that have similar vocabulary set. If we model comments of a spammer with a term-document matrix (rows as vocabularies and columns as documents), each columns will be similar to each other, and will become linearly dependent to each other. The rank of the term-document matrix will be \emph{relatively} lower than similar size matrix of non-spam user.

There are other ways to compute the similarity metric between columns such as cosine similarity, but traditional metrics usually defined in pairwise and not very intuitive to measure similarities of multiple documents as a whole. Average of all combinations of pairwise cosine similarities was suggested in \cite{Lim2010,Fei2013}, and it was found effected in detecting spammers. However, this type of approach is much slower than rank based metric since it needs to compute all possible combinations. Given a $n$ documents with $m$ vocabularies (term-document matrix is $\R^{m \times n}$), the average cosine measure needs $O(4m \cdot n^2)$. $4m$ is for a cosine similarity between two documents (note that this will be much slower in sparse vector multiplications), and $n^2$ is for all possible pairs. It is indeed slower than computing the structural rank $O(\tau + n)$. See section~\ref{sec:computation} for empirical computation results.

We propose to use structural rank (Section~\ref{sec:structural-rank}) for computing content similarity of a set of documents. 1) Solely considering the non-zero pattern will be enough to measure the content similarity of a set of documents. 2) Term-document matrices are usually very sparse and our case will be even more sparser as we deals with very short documents (comments). Bipartite graph traverse algorithm will be extremely efficient in this case. 3) It will be also much faster than other pairwise based similarity metrics.

\subsection{Spammer Score}

We propose this $SpammerScore$, which will be use to determine a spammer:
\begin{align}
  SpammerScore(A) &= 1 - \frac{StructuralRank( D(A) )}{N} \label{eq:score}\\
  & \quad \text{where $D(A)$ is a $M \times N$ term-document matrix of author $A$.}\notag
\end{align}

Higher the score, the relative structural rank will be lower, and will be determined as a spammer. For example, if one author keep posting the same contents over and over the score will be $1-\frac{1}{N}$. The other end will be $0$ when an author posted very different postings at each time: $1-\frac{N}{N} = 0$.

It is noteworthy to mention that our method can also be combined with other types of features such as spam dictionary or user profiles. Our method can be a good add-on features on spam detection systems providing a natural concept of duplicated comments. For example, our method is a good surrogate of average cosine similarity in \cite{Lim2010,Fei2013}.

\section{Experiment}

\subsection{Dataset}

We used NBA dataset for XDATA 2014 challenge\footnote{\url{http://www.darpa.mil/OpenCatalog/XDATA.html}}. The dataset contains 352936 comments from Yahoo sport and ESPN sport website. Comments were from public audience and were responses for NBA games of season 2011-2012$\sim$2013-2014 (3481 games). There were 42382 authors and most authors doesn't post much (35K authors have less than 10 comments). Standard preprocessing was performed to generate term-document matrices. We removed all HTML tags, lowercased, tokenized, stemmed, and removed stop words.

\subsection{Observations}

\begin{figure}
  \begin{center}
  \includegraphics[width=.65\linewidth]{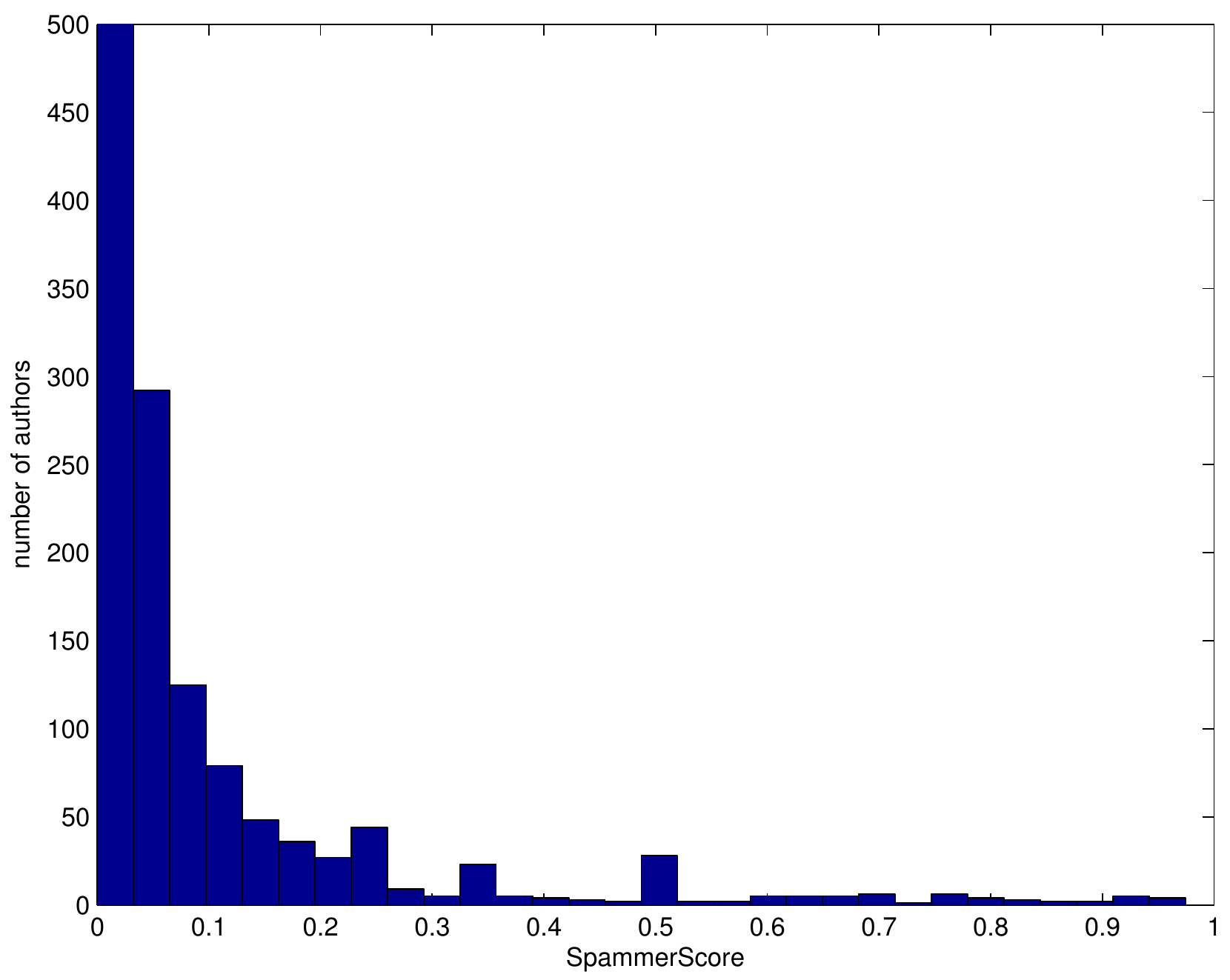}
  \end{center}
  \caption{Histogram of the $SpammerScore$ \eqref{eq:score} of 42K authors. Please note that upper part of histogram (around $SpammerScore=0$) are clipped for visualization. See text for details.}
  \label{fig:histogram}
\end{figure}

Figure~\ref{fig:histogram} shows histogram of 42K authors of our dataset. Most of author have $SpammerScore$ of $0$, but there are also many authors with non-zero $SpammerScore$ (1237 authors). Although, the sketchy authors are in small number, they produced a large portion of comments (28\% of entire comments). They averagely wrote 80.93 comments per author while other innocent authors averagely have 6.14 comments per person. It is not surprising fact since many spammers tent to post far more actively.

Belows are examples of comments from a sample author in each score range. Please note each quote indicates one comment.

\begin{itemize}
\item \textbf{0.1 $\sim$ 0.2}: author\#74 ($N=13,StructuralRank=11,SpammerScore = 0.1538$)\vspace{0.5em}\\
``man bum hole ?",
``burn hell !",
``night lmao",
``merri christma wizard fan",
``wizard win",
``merri christma",
``dirti wizard wash ur bum",
``shawn kemp back ! ! ! ! ! ! ! ! ! ! ! ! !",
``. ! ! ! ! ! ! ! ! ! ! ! ! ! ! ! ! ! ! ! ! !",
``lol",
``lol",
``merri christma sun fan ! ! ! !",
``lol"
\item \textbf{0.2 $\sim$ 0.3}: author\#2038 ($N=28,StructuralRank=20,SpammerScore = 0.2857$)\vspace{0.5em}\\
``thunderup",
``thunderup ! ! ! westbrook",
``thunderup ! ! !",
``thunderup",
``okc babi roll !",
``damn ref ! call made mav !",
``blazer season",
``bet guy indianapoli game board complain leav baltimor .",
``thunderup",
``meh , tough loss bound lose eventu .",
``true rocket fan credit , optim , fake suddenli rocket fan pt win hilari",
``home",
``not-rocket fan houston",
``harden ball hog troll",
``thunderup",
``surg bit charact night , profession opinion .",
``bulk dive paint aggress game man .",
``lmao , guess consensu heat thunder final thunder round ... .",
``not-, clip fan .",
``thunderup houston .",
``westbrook alright !",
``thunderup",
``kd humbl",
``thunderup !",
``thunderup ! ! !",
``thunderup !",
``",
``cook shoot !"
\item \textbf{0.3 $\sim$ 0.4}: author\#1900 ($N=32,StructuralRank=22,SpammerScore = 0.3125$)\vspace{0.5em}\\
``thunder start win streak ! ! ! thunder !",
``? ?",
``thunder !",
``steven adam top player posit year . book",
``thunder ! ! ! ! !",
``beat percentag year tough task .",
``thunder ! tough game , thunder win ! !",
``! home . thunder ! ! !",
``lebron kd . absolut worthless",
``thunder ! ! ! ! !",
``thunder ! beaten game . hold fight ! okc !",
``time back track . thunder !",
``thunder ! ! ! !",
``okc ! !",
``thunder ! ! ! !",
``codi palmer not-win 'em . game !",
``okc ! ! ! !",
``thunder ! !",
``justin breakdown russel westbrook serg eject buzzer beater record west not-great . thunder team road record claim .",
``mad , bro ?",
``thunder ! !",
``thunder ! ! ! ! !",
``ayn bland point proven",
``lol",
``russel claim not-elit player .",
``mad , bro ?",
``tnt yea . mad , bro ...",
``thunder ! ! ! !",
``bench play great night . young team warrior ? gs la",
``thunder ! !",
``okc",
``okc . not-easi . thunder win ! !"
\item \textbf{0.4 $\sim$ 0.5}: author\#16625 ($N=24,StructuralRank=14,SpammerScore = 0.4167$)\vspace{0.5em}\\
``wiz",
``wizard",
``win game wizard team year",
``wizard win",
``wiz",
``blow",
``bout wizard",
``row tonight wiz",
``wiz",
``john wall good",
``wizard",
``wizard",
``easi win wizard",
``good win wizard",
``wiz",
``wiz",
``wizard win game",
``wiz",
``wiz nice comeback",
``win wiz",
``wiz babi",
``wizard win today",
``wizard win",
``wiz"
\item \textbf{0.5 $\sim$ 0.6}: author\#18054 ($N=7,StructuralRank=3,SpammerScore = 0.5714$)\vspace{0.5em}\\
``random nbsp ;",
``random nbsp ; random nbsp ; random nbsp ; random nbsp ; random nbsp ;",
``random nbsp ; random nbsp ; random nbsp ; random nbsp ;",
``random nbsp ; random nbsp ; random nbsp ;",
``random nbsp ;",
``random nbsp ;",
``random nbsp ;"
\item \textbf{0.6 $\sim$ 0.7}: author\#17205 ($N=33,StructuralRank=12,SpammerScore = 0.6364$)\vspace{0.5em}\\
``forward buck play decent team . win streak nice play ? !",
``prost !",
``prost !",
``prost !",
``prost !",
``prost !",
``prost !",
``prost !",
``prost !",
``prost !",
``prost !",
``prost !",
``scott skile bo ryan nba ... ... flourish . rest team grow tire skile yell defens orient game , leuer sir .",
``prost !",
``ersan !",
``put gooden !",
``prost !",
``play leuer !",
``milwauke buck : fun team watch ? ! awhil !",
``put leuer !",
``buck play ? ?",
``prost !",
``prost !",
``prost !",
``prost !",
``prost !",
``prost !",
``prost !",
``prost !",
``play leuer !",
``prost !",
``prost !",
``ersan final decid join team !"
\item \textbf{0.7 $\sim$ 0.8}: author\#13938 ($N=7,StructuralRank=2,SpammerScore = 0.7143$)\vspace{0.5em}\\
``http :",
``http :",
``http :",
``http :",
``http :",
``http :",
``http :"
\item \textbf{0.8 $\sim$ 0.9}: author\#11049 ($N=41,StructuralRank=7,SpammerScore = 0.8293$)\vspace{0.5em}\\
``hate gg rz smell ballz",
``hate gg rz smell ballz",
``hate gg rz smell ballz",
``hate gg rz smell ballz",
``hate gg rz smell ballz",
``hate gg rz smell ballz",
``hate gg rz smell ballz",
``hate gg rz smell ballz",
``hate gg rz smell ballz",
``hate gg rz smell ballz",
``hate gg rz smell ballz",
``hate gg rz smell ballz",
``hate gg rz smell ballz",
``hate gg rz smell ballz",
``hate gg rz smell ballz",
``hate gg rz smell ballz",
``hate gg rz smell ballz",
``hate gg rz smell ballz",
``hate gg rz smell ballz",
``hate gg rz smell ballz",
``hate gg rz smell ballz",
``hate gg rz smell ballz",
``hate gg rz smell ballz",
``hate gg rz smell ballz",
``hate gg rz smell ballz",
``hate gg rz smell ballz",
``hate gg rz smell ballz",
``hate gg rz smell ballz",
``hate gg rz smell ballz",
``hate gg rz smell ballz",
``! scum bag , seek doctor",
``mom",
``hate gg rz smell ballz",
``hate gg rz smell ballz",
``hate gg rz smell ballz",
``hate gg rz smell ballz",
``hate gg rz smell ballz",
``hate gg rz smell ballz",
``hate gg rz smell ballz",
``hate gg rz smell ballz",
``hate gg rz smell ballz"
\item \textbf{0.9 $\sim$ 1.0}: author\#7383 ($N=52,StructuralRank=5,SpammerScore = 0.9038$)\vspace{0.5em}\\
``weaker wade",
``bad deal foy crawford",
``sd sport curs",
``sam cassel",
``dsfaasdfasdf",
``dsfaasdfasdf",
``dsfaasdfasdf",
``dsfaasdfasdf",
``dsfaasdfasdf",
``dsfaasdfasdf",
``dsfaasdfasdf",
``dsfaasdfasdf",
``dsfaasdfasdf",
``dsfaasdfasdf",
``dsfaasdfasdf",
``dsfaasdfasdf",
``dsfaasdfasdf",
``dsfaasdfasdf",
``dsfaasdfasdf",
``dsfaasdfasdf",
``dsfaasdfasdf",
``dsfaasdfasdf",
``dsfaasdfasdf",
``dsfaasdfasdf",
``dsfaasdfasdf",
``dsfaasdfasdf",
``dsfaasdfasdf",
``dsfaasdfasdf",
``dsfaasdfasdf",
``dsfaasdfasdf",
``dsfaasdfasdf",
``dsfaasdfasdf",
``dsfaasdfasdf",
``dsfaasdfasdf",
``dsfaasdfasdf",
``dsfaasdfasdf",
``dsfaasdfasdf",
``dsfaasdfasdf",
``dsfaasdfasdf",
``dsfaasdfasdf",
``dsfaasdfasdf",
``dsfaasdfasdf",
``dsfaasdfasdf",
``dsfaasdfasdf",
``dsfaasdfasdf",
``dsfaasdfasdf",
``dsfaasdfasdf",
``dsfaasdfasdf",
``dsfaasdfasdf",
``dsfaasdfasdf",
``dsfaasdfasdf",
``dsfaasdfasdf"
\end{itemize}

\vspace{1em}
As we expected, the higher $SpammerScore$ is, the larger number of similar contents an author posts. Choosing a right threshold to determine spammers is not trivial. We may augment our $SpammerScore$ feature with other features and learn a classifier on a labeled training set to automatically determine the threshold.

\subsection{Computation}\label{sec:computation}

\begin{table}
\centering
\begin{tabular}{l r}
\toprule
\bf Method & \bf Time (sec) \\
\midrule
Structural rank (our proposal)  &     1.523 \\
Rank (using sparse SVD)         &   537.630 \\
Sparse numeric rank             &   265.614 \\
Average cosine similarity       & 21435.117 \\
\bottomrule
\end{tabular}
\caption{Time comparison of four methods to compute content similarity of 42K term-document matrices. See text for details.}\label{tbl:speed}
\end{table}

In section~\ref{sec:method}, we claimed that we chose structural rank for faster content similarity computation. In this section, we empirically compare the computation time with other comparable methods as it will be a bottleneck of computing the $SpammerScore$. The task involve 42K term-document matrices with average size of $7368 \times 11$ (dictionary size is 7346 and average number of comments per author is 11).

Table~\ref{tbl:speed} includes time consumptions of four different content similarity methods. 1) The structural rank is computed with Matlab's implementation \cite{Pothen1990}, and is our proposal. 2) We also computed a true rank of the matrix using sparse SVD, which is one of the most accurate way to determine matrix rank. The method is very close to default Matlab procedure, but we change dense SVD to sparse SVD as the default method failed to run. 3) There is another method that directly compute numeric rank of a sparse matrix without SVD. We uses a \emph{SPNRANK}\footnote{Leslie Foster: \url{http://www.math.sjsu.edu/singular/matrices/software/SJsingular/spnrank.m}} for this. 4) The average cosine similarity \cite{Lim2010,Fei2013} was computed by Matlab implementation \emph{PDIST}\footnote{http://www.mathworks.com/help/stats/pdist.html} then take average of all pairwise similarities.

Results in Table~\ref{tbl:speed} show that the proposed method is far faster than any other method in large magnitude. It is because our dataset is extremely sparse (average density = 0.001206). Please note that it is a usual case of handling comments as lengths of document is very short. We believe this efficient computation is especially useful in practice since we usually have billions of users and comments.

\newpage
\section{Discussion}

In this paper, we introduced an efficient way detecting spammers by structural rank of per-author term-document matrix. The use of structural rank turned out to be much faster than using traditional rank in our scenario. We hope to extend this line of work to include richer set of features to achieve state-of-the-art spammer detection performance.

We also feel the structural rank needs an additional attention measuring multi-document similarity; for example, evaluating document clusters, document relevance, or in search engines.

\bibliographystyle{plain}
\bibliography{../../share/groupPapers,../../share/externalPapers}

\end{document}